\documentclass[a4paper,fleqn,usenatbib]{mnras}

\usepackage{multicol}
\usepackage[T1]{fontenc}
\usepackage{ae,aecompl}

\usepackage{graphicx}	
\usepackage{amsmath}	
\usepackage{amssymb}	

\title[Flickering in Kepler dwarf novae]
{Flickering around the outburst cycle in Kepler dwarf novae}

\author[A. Bruch]{Albert Bruch
\\
Laborat\'orio Nacional de Astrof\'{\i}sica, Rua Estados Unidos, 154, 
CEP 37500-364, Itajub\'a, MG, Brazil
}

\date{Accepted XXX. Received YYY; in original form ZZZ}

\pubyear{2019}

\begin{document}
\label{firstpage}
\pagerange{\pageref{firstpage}--\pageref{lastpage}}
\maketitle

\begin{abstract}
Taking advantage of 
the unparallel quantity and quality of high cadence Kepler light 
curves of several dwarf novae, the strength of the flickering and the 
high frequency spectral index of their power spectra are investigated as
a function of magnitude around the outburst cycle of these systems. 
Previous work suggesting that the flickering strength (on a magnitude
scale) is practically constant above a given brightness threshold and only
rises at fainter magnitudes is confirmed for most of the investigated systems.
As a new feature, a hysteresis in the flickering strength is seen in the sense
that at the same magnitude level flickering is stronger during decline from
outburst than during the rise. A similar hysteresis is also seen in the 
spectral index. In both cases, it can qualitatively be explained under 
plausible assumptions within the DIM model for dwarf nova outbursts. 
\end{abstract}




\section{Introduction}
\label{Introduction}

The phenomenon of flickering is ubiquitous in astronomical systems where 
accretion of matter onto a central object occurs. In the optical regime it 
is most conspicuous in cataclysmic variables (CVs) where it manifests itself 
as a continuous series of apparently stochastic overlapping flares which --
depending on the particular system and photometric state -- can reach full 
amplitudes between some tens of milli-magnitudes up to more than an entire 
magnitude. In spite of an increasing number of specific studies in recent
years, flickering is still one of the lesser understood properties of CVs.
For a review about the state of the art in this field see the introduction
to \citet{Bruch15}.

CVs are close binary stars with orbital periods between $\approx$80~min and
rarely more than about 10 hours, where a late type companion -- 
the secondary --, 
which is in most cases on or close to the main sequence, fills its Roche lobe 
and transfers matter to a white dwarf primary. In the absence of strong 
magnetic fields conservation of angular momentum forces this material to 
form an accretion disk around the white dwarf where viscous forces cause 
it to slowly move inwards and to finally settle on the surface of the 
compact star. In optical light this disk is almost always the most luminous 
part of the system. For a comprehensive description of most aspects of CVs, 
see, e.g., \citet{Warner95}. 

The most common of the many subtypes of CVs are the dwarf novae. They are
characterized by semi-regularly spaced outbursts with amplitudes ranging from
roughly 2 -- 8~mag, occurring at intervals which may be as short as a couple of 
days in some systems and as long as decades in others. They last a few days 
to rarely more than two weeks. In general, the dichotomy between quiescent 
states and outbursts is explained by a thermal instability which develops in 
the accretion disk \citep{Smak71, Osaki74, Lasota01}. In short, 
during quiescence the disk is in a low viscosity state such that less
matter is accreted onto the surface of the white dwarf than is transferred from
the secondary. Consequently, the column density (and the temperature) of the
disk increases, until a thermal instability caused by the ionization of
hydrogen sets in, resulting in an increase of the viscosity and thus an
increased dumping of matter onto the central object. The corresponding
release of potential energy manifests itself as an outburst. After most of
the transferred matter is drained from the accretion disk the system returns
to the quiescent state and the cycle starts anew. 

The non-trivial problem to quantify the strength of the flickering (or the 
flux of the flickering light source relative to the total light of the 
system) in an objective
way, enabling a comparison between different systems and photometric states,
has recently been addressed by \citet{Bruch21} (hereafter referred to as B21). 
He parameterized the flickering strength,
expressed in magnitudes, by the full width at half maximum (FWHM) of a 
Gaussian fit to the distribution of data points in a flickering light curve,
after applying corrections for various systematic effects (see below for
details). B21 called this quantity $A_{60}$, the number 60 indicating that
it only refers to flickering occurring on times scales below 60~min. He
measured $A_{60}$ in several thousand light curves of more than 100 CVs.

Another -- and more often applied -- approach to quantified the properties
of flickering is based on a frequency analysis, i.e., the study of the power
spectra of CV light curves 
\citep[e.g.,][]{Dobrotka10, Kraicheva99, Scaringi12}. Most of these efforts
concentrate on the indentification and 
characterization of break points in the power spectra 
in a double logarithmic representation, and less on their slope (spectral 
index) at high frequencies. 

The flickering amplitude in dwarf novae is observed to depend on the outburst 
state. It has long been known that flickering is much stronger during 
quiescence than during outbursts. But this has never been quantified
before B21 systematically studied the evolution of the flickering strength
around the dwarf nova outburst cycle. He found that, expressed in magnitudes,
flickering in dwarf nova has the tendency to remain constant whenever the
system is brighter than a threshold magnitude level which is approximately 
half way between quiescence and outburst maximum. Only below this level the 
amplitude rises rapidly with decreasing system brightness. This means that on 
the flux scale the brightness of the flickering light source is proportional 
to the total flux above the threshold, while its contribution to the total 
light increases when the system is fainter. Concerning the frequency behaviour,
the dependence of the spectral index on the outburst state has never been
studied.

The data used by B21 came from multiple terrestrial observatories. Their 
quality and suitability for flickering studies are rather inhomogeneous.
The above mentioned various corrections necessary to render the flickering 
strength measured in different light curves comparable to each other are 
frequently not known to a high precision and thus introduce uncertainties. 
Observations with sufficient time resolution around the outburst cycle of 
dwarf novae are most often quite fragmentary. Power spectra of individual 
light curves are in general rather noisy. All this makes it difficult to 
follow the evolution of the flickering strength and the spectral index 
around the outburst cycle of 
individual dwarf novae. This would only be different if high quality data 
sets covering various cycles were available.

Such data indeed exist for a few systems. Among the many stars observed by
the Kepler space mission \citep{Koch10} five dwarf novae were observed in
short cadence mode for up to almost four years, in some cases with only small
interruptions. This means that homogeneous light curves, continuous over long
time intervals, with a time resolution of $\approx$59~s are available, covering
numerous outbursts and quiescent phases. These data are used here to
investigate the evolution of the flickering strength and of the spectral
index from quiescence to outburst and back again. 

\begin{figure}
	\includegraphics[width=\columnwidth]{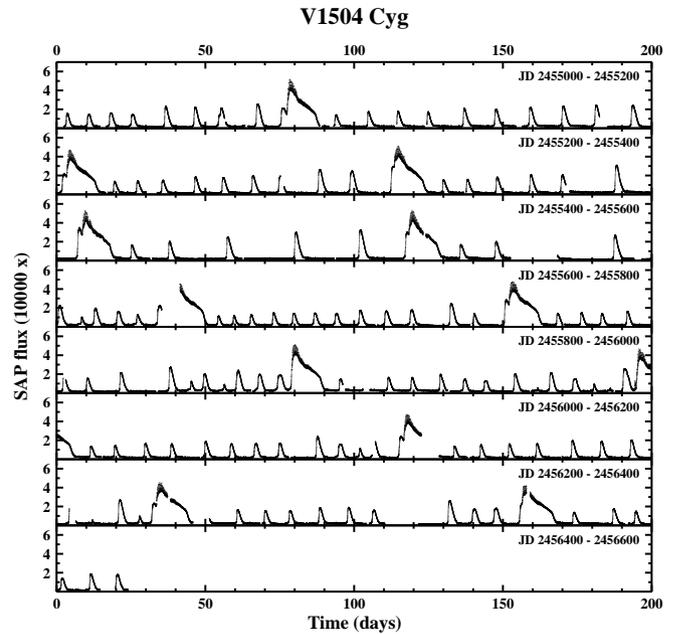}
      \caption[]{Long term light curve of V1504 Cyg. Each frame contains
                 a 200 day section of the entire data set. Julian Dates of
                 of start and end of each section are indicated in the upper 
                 right corners of the frames.}
\label{v1504cyg-longtlc}
\end{figure}

\begin{figure}
	\includegraphics[width=\columnwidth]{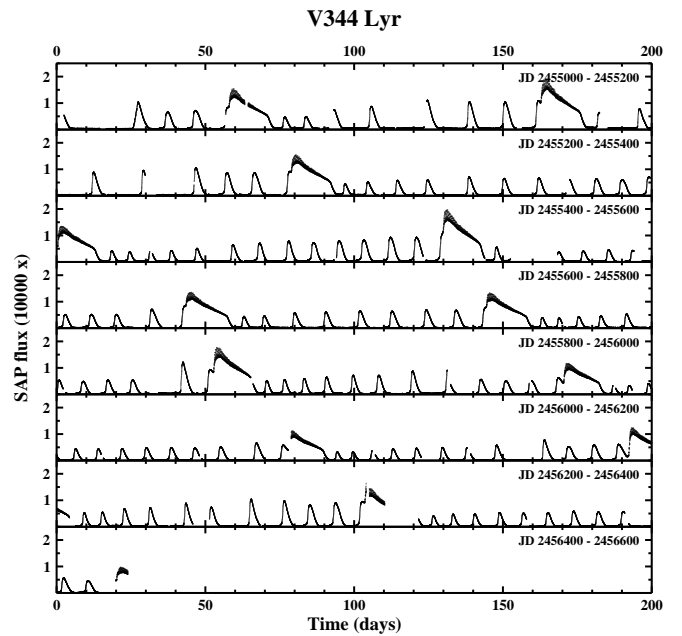}
      \caption[]{Same as Fig.~\ref{v1504cyg-longtlc} for V344~Lyr.}
\label{v344lyr-longtlc}
\end{figure}

\begin{figure}
	\includegraphics[width=\columnwidth]{v447lyr-longtlc.eps}
      \caption[]{Same as Fig.~\ref{v1504cyg-longtlc} for V447~Lyr.}
\label{v447lyr-longtlc}
\end{figure}

\begin{figure}
	\includegraphics[width=\columnwidth]{v516lyr-longtlc.eps}
      \caption[]{Same as Fig.~\ref{v1504cyg-longtlc} for V516~Lyr.}
\label{v516lyr-longtlc}
\end{figure}

\begin{figure}
	\includegraphics[width=\columnwidth]{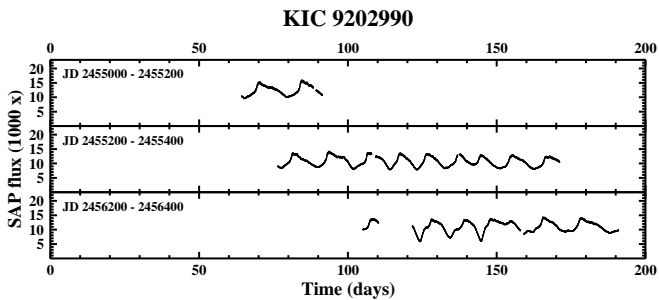}
      \caption[]{Same as Fig.~\ref{v1504cyg-longtlc} for KIC 9202990.}
\label{kic9202990-longtlc}
\end{figure}

\section{The target stars}
\label{The target stars}

The dwarf novae for which Kepler short
cadence data are available are V1504~Cyg, V344~Lyr, V447~Lyr, V516~Lyr
and KIC~9202990. For some of these targets, long cadence observations 
(time resolution: $\approx$29~min) are also
available. They are, however, not useful in the present context because
they cannot resolve the flickering activity and are thus ignored.

Both, V1504~Cyg and V344~Lyr are short period dwarf novae of the SU~UMa
subtype. This means, that apart from normal outbursts they occasionally 
exhibit superoutbursts which are significantly longer and brighter than
normal ones. These systems were discovered by \citet{Beljawski36} and 
\citet{Hoffmeister66}, respectively. Their Kepler data have 
repeatedly been analysed in the literature, often in the same publication.
These concentrate on studies of the outburst behaviour and of (positive and
negative) superhumps \citep{Still10, Wood11, Cannizzo12, Osaki13a, Osaki13b,
Osaki14}. The investigation of the flickering behaviour in these stars is
restricted to an analysis of the frequency spectrum by \citet{Dobrotka15}
and \citet{Dobrotka16}.

Compared to V1504~Cyg and V344~Lyr, much less is known about the other
dwarf novae regarded here. Except for KIC~9202990 they are also significantly 
fainter which turns
the secure measurement of their flickering properties more difficult, in
particular in quiescence.

V516~Cyg is another SU~UMa type dwarf nova. It was discovered as a variable
star by \citet{Kaluzny92} and confirmed as 
a cataclysmic variable by \citet{Kaluzny97}. A detailed study based on Kepler
data was performed by \citet{Kato13} who studied mainly the outburst behaviour.
They draw attention to the occasional presence of double outbursts.
 
V447~Lyr was discovered as a variable star by \citet{Romano72}. Its orbital 
period lies above the range of periods of SU~UMa stars. The only detailed 
study of this system was published by \citet{Ramsay12} 
who combined optical spectroscopy with Kepler light curves. V447~Lyr is 
eclipsing at a period of 0.1556270~d. Individual eclipses
can only be securely identified during outburst, while in quiescence they
are drowned in noise. No further periods were detected in the Kepler data.
V447~Lyr exhibits long and short outbursts. It is
remarkable that in all long outbursts the initial rise is followed
by a small decline before the final rise to maximum. As was pointed out
by \citet{Ramsay12} this resembles very much
the structure of superoutbursts in SU~UMa stars which are often triggered by
a normal outburst. 

KIC~9202990 is unusual in the sense that its Kepler light curve is 
characterized by a continuous series of low amplitude modulations repeating 
on the time scale of two weeks \citep{Ostensen10}. Superposed is a
shorter stable period of $\approx$4~h.  \citet{Ramsay16}
discuss these observations in more detail and interpret the
longer (quasi-) periodicity as stunted outbursts as sometimes seen in 
dwarf novae (but never as the continuous series seen in KIC9202990). The
4~h period is considered to be orbital. 

\section{The data}
\label{The data}

All data used in this study were retrieved from the Barbara A.\ Mikulski
Archive for Space Telescopes 
(MAST)\footnote{https://mast.stsci.edu/portal/Mashub/clients/MAST/Portal.html}.
Among the various formats in which they are available I chose the SAP (Simple
Aperture Photometry) flux data\footnote{SAP data are preferred here 
over PDC-SAP 
(Presearch Data Conditioning Module Simple Aperture Photometry fluxes) 
data because the latter contain intervals with an anomalous flickering
behaviour (e.g., BJD 2455415-31 in V344~Lyr) which is not observed in the
SAP flux data. These intervals invariably coincide with monthly datasets
which contain a superoutburst and end with the beginning of the next dataset. 
It is therefore close at hand to suspect that
the unusual flickering is an artefact of data reduction.}
restricting myself to data obtained in short cadence mode.
Some among the millions of data points
were identified to deviate strongly from neighbouring points in the 
light curves (more so for the fainter objects than for the brighter ones).
These must be considered erroneous. Consequently they have been removed. 
The SAP flux of V516~Lyr in the first two monthly Keper data sets 
(BJD 2455568-636) is systematically lower (by about 1.3~mag in quiescence) 
than in all other data sets. They have therefore been ignored. 

The long term Kepler light curves which have never been shown in
their entirety are reproduced in Figs.~\ref{v1504cyg-longtlc} 
- \ref{kic9202990-longtlc} which are organized such that each panel contains 
a 200 day section. 

\section{The strength of the flickering}
\label{The strength of the flickering}

Flickering manifests itself as an (at least apparently) stochastic
process. The problem of how to measure its strength has been addressed
in detail by B21. Here, I follow the prescriptions outlined
in that publication. As already briefly mentioned in Sect.~\ref{Introduction}, 
the flickering strength is parameterized by the FWHM $A_{60}$ of a Gaussian 
fit to the distribution of data points in a flickering light curve (expressed
in magnitudes) from which variations on time scales longer than
the typical flickering time scales (here chosen to be 60~min) have been 
subtracted. In order to render this quantity comparable for different 
systems and photometric states, some conditions have to be met and several 
corrections have to be applied as listed in Sect.~2 of B21. Due to the 
homogeneous characteristics of the Kepler data, the latter are either much 
easier to quantify than in more inhomogeneous data typical for terrestrial 
observations, or can even be ignored.

In order to follow the variations of the flickering strength around the 
outburst cycle of the dwarf novae their Kepler 
light curves of are cut into sections of
8 hours (about the upper limit of a long light curves observable in
terrestrial observations). This is long enough for the flickering in each
section to meet the requirement to be representative.

The second requirement, i.e.,
separating variations unrelated to flickering, is met by subtracting a
\citet{Savitzky64} filtered light curve with a cut-off time scale of 60~min
from the original one. This effectively removes orbital 
variations as well as modulations on longer time scales. Thus, only flickering
occurring on times scales of less than 60~min are regarded (as has been
done by B21). It turned out that this procedure could sometimes not remove
satisfactorily the pointed maxima of superhump variations occuring during the 
supermaxima of the SU~UMa stars. Therefore, supermaxima (and subsequent
time intervals with persisting superhumps identified in power spectra
of the data in these intervals) were disregarded. 

A correction for data noise, difficult in the general
case because the noise level is often not well known, is easy here because
the original Kepler data files provide the flux error for each data point of 
the light curve. This is used to correct the measured FWHM of the data
distribution function for the effect of noise. 

A reduction of $A_{60}$ to a standard time resolution of the data (and thus
correcting for the effect that a finite integration time always acts as a
filter, reducing the difference between local minima and maxima in a 
flickering light curve) is only required if the flickering in light curves 
obtained using different integration times is to be compared. However, the 
Kepler short cadence data were all observed with the same time resolution, 
rendering a corresponding correction unnecessary.\footnote{For the record: 
In order to reduce the flickering strength to the standard time resolution of
5~s defined by B21 the measured and noise corrected FWHM should be divided by
0.85.}

Applying a correction for the contribution of the secondary star is not
necessary for the SU~UMa stars of the current sample because that contribution
is negligible in short period cataclysmic variables. This may be different
for V447~Lyr and KIC~9202990 which have longer orbital periods. No 
quantitative information for the secondary star contribution in these 
systems is available. However, since no comparison of their flickering
strength with that in other CVs will be performed, neglecting a corresponding
correction does not invalidate the present results.

\begin{figure*}
	\includegraphics[width=\textwidth]{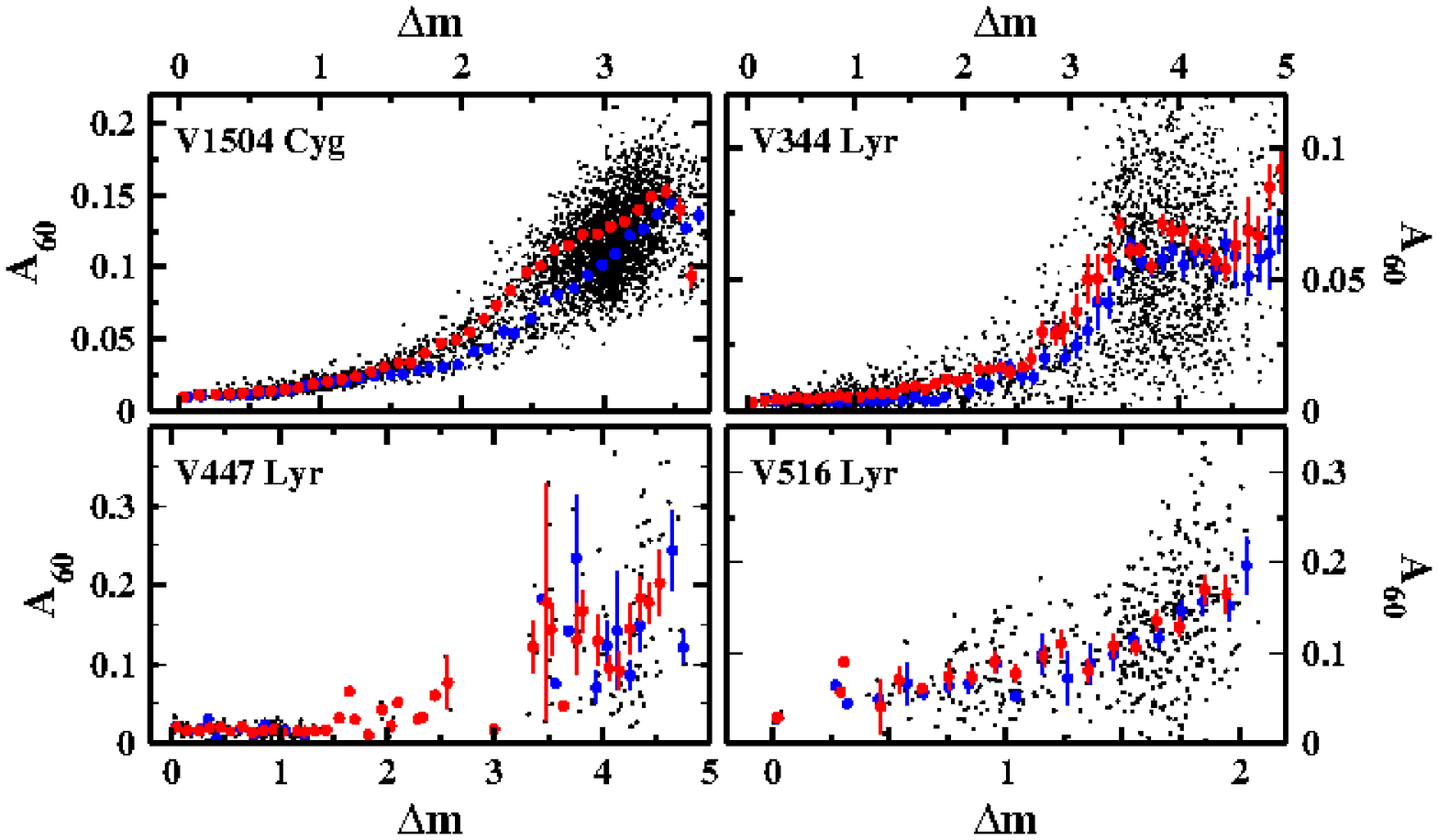}
      \caption[]{The flickering strength $A_{60}$ (small black dots) measured
      in individual light curves sections of four dwarf novae as a function of
      magnitude. The zero point of the magnitude scale corresponds to the
      brightness maximum of the system in the outburst cycle. The larger dots
      are the averages of the individual $A_{60}$ values in intervals
      of 0.1~mag measured before (blue) and after (red) outburst maximum.}
\label{flickamp}
\end{figure*}

Fig.~\ref{flickamp} shows the results for four of the five dwarf novae
studied here. In all cases the small black dots represent $A_{60}$ measured
in the individual 8~h sections of the light curve, plotted against the
average magnitude of all data points in that section. It turned out that 
excessive noise in the fainter systems V447~Lyr and V516~Lyr (but also to
some extend in V344~Lyr) inhibited to measure $A_{60}$ in some sections 
which were thus ignored. The magnitude scale is arbitrary and has been 
chosen such that the zero point corresponds to the brightest average 
magnitude in any of the light curve sections (and thus to the
brightest maximum) of each dwarf nova. The 
coloured dots are average values of $A_{60}$, binned in intervals of 
$\Delta m = 0.1$. The blue ones refer to $A_{60}$ measured in sections 
before outburst maxima (from quiescence halfway between a given outburst 
maximum and the preceding one), the red ones after outburst (from outburst
maximum to quiescence halfway to the next outburst). The error bars (mostly
smaller than the plot symbols for V1504~Cyg) represent the standard error
of the mean.

Two interesting features can be seen in the results. The quality of the data
of B21 permitted him to see only on the average over several systems that
flickering remains constant on the magnitude scale whenever a dwarf nova
system remains above a certain brightness level during the outburst cycle
and starts to grow only when it is closer to quiescence. Here, this is 
confirmed in individual systems in at least two of the four investigated 
cases. In V344~Lyr $A_{60}$ is quite constant as long as the system is less
than $\approx$1-2~mag below maximum [more so before maximum (blue dots) than
afterwards (red dots)]. Thereafter, first a gradual and then a
faster rise sets in. In V447~Lyr the $A_{60} - \Delta{\rm m}$ relationship
remains absolutely flat up to 1.5~mag below maximum before a gradual rise
begins. The case of V516~Lyr is less clear. But disregarding the lone 
point close to $\Delta m = 0$, the distribution of data points appears also
to be compatible with a constant close to outburst maximum. In contrast,
V1504~Cyg does not show such a plateau phase, but $A_{60}$ continually drops
as the star gets brighter. However, even in this case a change
in slope occurs close to $\Delta m \approx 1.9$ as becomes obvious
when the average difference in $A_{60}$ between neighbouring datapoints is 
regarded. 

While this behaviour thus confirms the results of B21, the second feature
reveiled here has never been seen before. Most clearly in
V1504~Cyg, but also in V344~Lyr there is a hysteresis in the flickering
strength around the outburst cycle. On the rise to outburst (blue dots in
Fig.~\ref{flickamp}) $A_{60}$ is systematically smaller at a given magnitude
than on the decline (red dots). This is not seen in V447~Lyr and V516~Lyr,
either because the hysteresis is really absent, or because the inferior
quality of the respective light curves inhibits its detection.

\begin{figure}
	\includegraphics[width=\columnwidth]{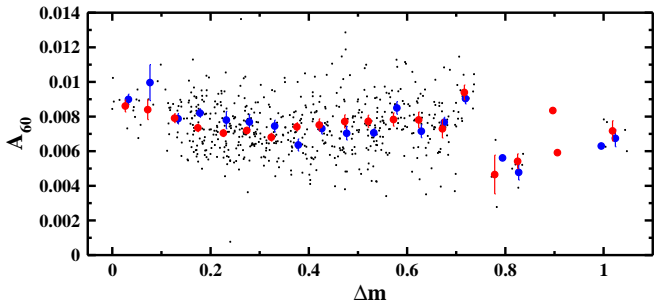}
      \caption[]{Same as Fig.~\ref{flickamp} for KIC~9202990.}
\label{flickamp-kic}
\end{figure}

In the unusual system KIC~9202990 the flickering strength behaves differently 
(Fig.~\ref{flickamp-kic}). Over the quite limited range of variability
flickering remains on a approximately constant
low level of $A_{60} = 0.0074 \pm 0.0016$~mag with possibly a slight
increase during the mimima as well as the maxima of the outburst cycle.
There may be a slight decrease of $A_{60}$ during the exceptionally 
faint minima ($\Delta m > 0.75$) observed in the interval JD~2456320--50
(see Fig.~\ref{kic9202990-longtlc}). But statistics are very poor in this
magnitude range.

In their investigation of flickering in the Kepler data of V1504~Cyg
\citet{Dobrotka15}, among other aspects, study the dependence of the 
scatter of the data in small time intervals 
as a function of flux during quiescence and outburst maximum. 
This can be considered as a proxy of the flickering strength. 
This relation is normally referred to as the rms-flux relation. This
denomination is, however, misleading because it does not the relate
the rms (root mean square), but the standard deviation of the sample of
considered data points to the flux \citep[see the definition 
of $\sigma_{\rm rms}$ according to eq.~1 of][]{Dobrotka15}.
Adopting the established, although erroneous, nomenclature this definition
renders the rms as a measure ot the flickering strength conceptually quite
similar to $A_{60}$ as defined by B21. In both cases, the distribution of
data points in a light curve interval is regarded. In fact, calculating
$\sigma_{\rm rms}$ for V1504~Cyg and V344~Lyr [using intervals comprising 
10 data points as has been
done by \citet{Dobrotka15}], multiplying the results by the ratio
between the standard deviation and the full width at half maximum of a 
Gaussian to account for the different definitions of $\sigma_{\rm rms}$ and
$A_{60}$, and transforming fluxes into magnitudes yields a relation
just as shown in Fig.~\ref{flickamp}. 

\begin{figure}
	\includegraphics[width=\columnwidth]{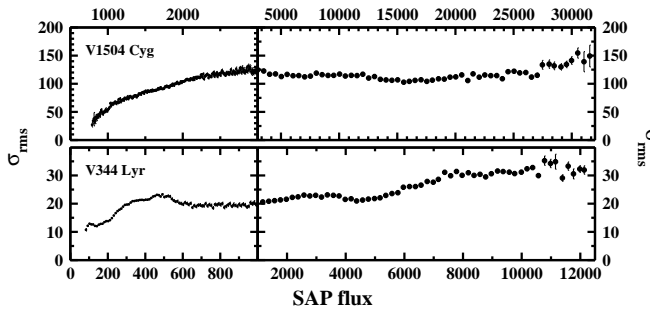}
      \caption[]{The rms-flux relation for V1504~Cyg and V344~Lyr. For
                 visualization purposes different scales for the
                 lower and higher flux regimes are used in the left
                 and right hand frames.}
\label{rms-flux-rel}
\end{figure}

The corresponding relations on the flux scale are shown in 
Fig.~\ref{rms-flux-rel} [using different resolutions for the low (left) and
high (right) flux regimes for better visualization]. For V1504~Cyg, the 
results of \citet{Dobrotka15} are recovered, most importantly the linear
$\sigma_{\rm rms}$ -- flux relation during quiescence which does not
extend to outburst phases. However, at low fluxes [not covered by 
\citet{Dobrotka15}] a significant steepening is observed. In V344~Lyr,
a linear $\sigma_{\rm rms}$ -- flux relation is present at most over a 
quite limited flux range. One reason may be the dominance of noise in the
low flux regime which causes a flattening of the relation. The slight
hump at intermediate fluxes (best seen in V344~Lyr between 2000 and 4000
SAP flux units) is not due to flickering  
but reflects an increased standard deviation
of the data points in particular during the steep rise to maximum because
of a significant increase in brightness even over the short time scale 
comprised by the 10 contributing data points. At high fluxes, the rise 
of $\sigma_{\rm rms}$
is steeper in V344~Lyr than in V1504~Cyg, reflecting the better expressed
constant phase of the flickering strength on the magnitude scale until well
below maximum in V344~Lyr as compared to V1504~Cyg (see Fig.~\ref{flickamp}).

A distinctive feature in the rms-flux relation of V344~Lyr is a hump
centred on 500 SAP flux units which interrupts the otherwise almost 
flat relation above the upper end of the linear rms-flux rise. A similar
structure was also observed by \citet{Dobrotka16} (their Fig.~2) who 
interpreted the rising part of the hump as a continuation of the linear 
rms-flux relation observed at lower levels and explain the delay by the
contribution of non-flickering flux provided by the light source of a
negative superhump. Considering that the intervals with positive 
superhump have been excluded from the present study 
(Sect.~\ref{The strength of the flickering}), that
no similar structure is seen in the case of V1504~Cyg, and that
negative superhumps only persist for limited time intervals in V344~Lyr and
V1504~Cyg \citep{Osaki14}, it is questionable if superhumps can cause
such a prominent feature in the rms-flux curve of V344~Lyr based on
data comprising predominantely intervals without superhumps. The flux 
level of $\approx$ 500 SAP units in V344~Lyr already corresponds to the early 
rise to and the late decline from outbursts. The hump may therefore be 
related to changing flickering properties during either or both of these 
transition phases.

\section{The spectral index of the flickering}
\label{The spectral index of the flickering}

Since the pioneering studies
of \citet{Elsworth82} it is well known that at high frequencies flickering 
manifests itself as red noise in the power spectra of cataclysmic variable 
light curves, meaning that $P \propto f^{-\gamma}$, where 
$P$ is the power, $f$ the frequency and $\gamma$ the spectral index. At very 
high frequencies white noise due to (approximately) Gaussian measurement 
errors takes over. On the double logarithmic scale, red noise causes a 
linear drop of the power with increasing frequency with a slope of $-\gamma$, 
and white noise results in a constant power level. Here, I will investigate
if and in which way $\gamma$ changes over the dwarf nova outburst cycle. 

\begin{figure}
	\includegraphics[width=\columnwidth]{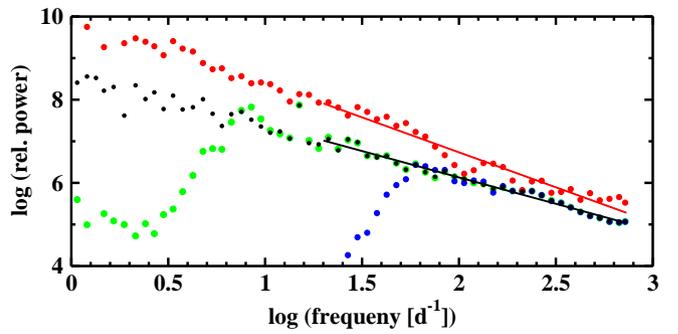}
      \caption[]{Power spectrum of a 15~day quiescent section of the light
                curve of V1504~Cyg on a double logarithmic
                scale, binned in intervals of $\Delta \log{f} = 0.05$
                (black dots). The red dots represent the power spectrum
                of the same data after adding a gradient typical for the
                decline of V1504~Cyg from outburst maximum. The black and
                red solid lines are linear least square fits to the respective
                data points between $f = 20$~d$^{-1}$ and the Nyquist 
                frequency. The blue and green dots are the power spectra of
                the light curve (including gradient) after removing variations
                on time scales of 1 and 8 hours, respectively.} 
\label{spec-index-method}
\end{figure}

As in the case of the flickering strength, care must be taken to avoid that
variations on longer time scales introduce a bias. This holds true particularly
during the rise and decline phases of the outbursts. Even if long time
scales correspond to low frequencies such variations have a significant
bearing also on the high frequency part of the power spectra. This can be
seen in Fig.~\ref{spec-index-method} where the black dots represent the 
double logarithmic power spectrum (Lomb-Scargle periodogram 
\citep{Lomb76, Scargle82}, using the normalization of 
\citet{Horne86}) binned in
intervals of 0.05 in $\log(f)$ 
(where $f$ is the frequencies expressed in cycles/day), of a particularly
long ($\approx$15 days) quiescent phase of V1504~Cyg. It exhibits excellently 
the exponential drop of power towards high frequencies up to the Nyquist 
frequency, as is shown by the black solid line which is a least squares fit 
to the data points above $f=20$~d$^{-1}$ (thus avoiding lower frequencies which
may be influenced by orbital variations; see below). The red dots represent 
the power spectrum of the same data, after a gradient typical of the outburst 
decline phase of V1504~Cyg has been added. This additional slope in 
the data changes the power spectrum in two ways: The spectral index becomes 
steeper, and some wiggles appear at high 
frequencies. The impact of the gradient should be even larger if typical
values for the much steeper rise to outburst maximum were adopted. 

Both effects can be avoided by subtracting a filtered version from the light 
curve (including the gradient). 
Using for this purpose a Savitzky-Golay filter with a cut-off time
scale of 60~min as used to measure the flickering strength results in the 
power spectrum represented by the blue dots in the figure. Now, the power
at low frequencies is strongly reduced, but at high frequencies it follows
very well the power spectrum of the original data. However, the turn-over
is at frequencies well above 20~d$^{-1}$, reducing significantly the frequency
range suitable to measure $\gamma$. This unduly reduces the precision of
$\gamma$ in particular when smaller light curve sections (fewer data) are 
contemplated and when white noise does not permit to use all frequencies
up to the Nyquist limit to measure $\gamma$. Choosing 8 hours for the filter 
cut-off time scale (i.e., the time base of the light curve sections used to
measure the flickering strength) results in the green dots in 
Fig.~\ref{spec-index-method}. They also follow excellently the power spectrum
of the original data over the entire frequency range suitable to measure
$\gamma$. 

\begin{figure}
	\includegraphics[width=\columnwidth]{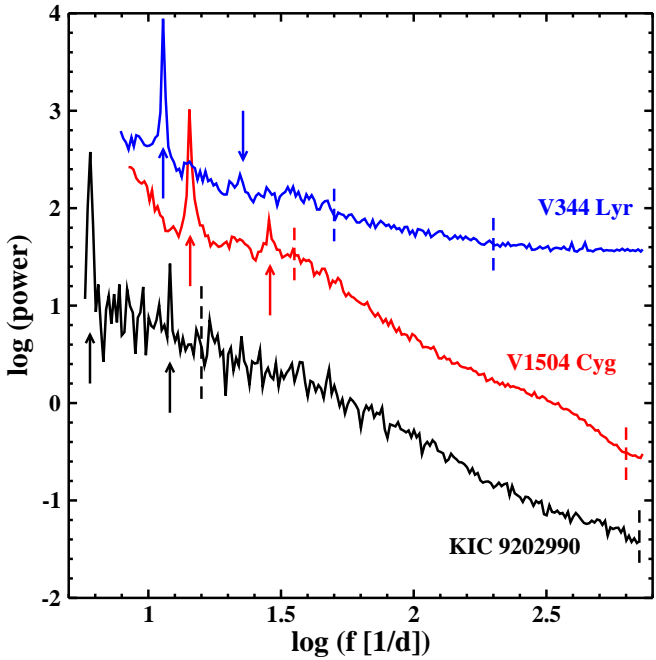}
      \caption[]{Average power spectra of V1504~Cyg, V344~Lyr (only
                 quiescent phases) and KIC~9202990 (entire light curve),
                 binned in intervals of 0.01 in log(frequency). The upper
                 frequency limit corresponds to the Nyquist frequency of
                 the data. Peaks
                 caused by orbital variations and their first harmonics
                 are marked by arrows. The broken vertical lines indicate
                 the range of frequencies used to measure the spectral
                 index. (For clarity, the graphs for V1504~Cyg and V344~Lyr 
                 are shifted vertically by an arbitrary amount.)} 
\label{powersp}
\end{figure}

I restrict the frequency analysis to V1504~Cyg, V344~Lyr and KIC~9202990,
because the data of V447~Lyr and V516~Lyr are so noisy that white noise
already dominates the power spectra at comparatively low frequencies. I
first regard the average (double logarithmic) power spectrum of the three 
stars (Fig.~\ref{powersp}), based only on the quiescent phases of V1504~Cyg 
and V344~Lyr (the continuous small amplitude variations of KIC~9202990 do 
not warrant to make a distinction between quiescence and outburst in this
system).

In all cases the orbital periods and their first harmonics manifest themselves
as well expressed peaks in the power spectra (marked by arrows in 
Fig.~\ref{powersp}). Avoiding these frequency ranges, the spectral index
is measured in the intervals delimited by the broken vertical lines. The
power spectrum of KIC~9202990 is declining very much linearly above 
$\log(f [{\rm d^{-1}}]) = 1.2$ up to the Nyquist frequency. With a slightly
steeper spectral index, the average power spectrum of V1504~Cyg shows a 
slight deviation from linearity (i.e, a small shoulder close to   
$\log(f [{\rm d^{-1}}]) = 2.5$) and an indication of a turn-over to white
noise close to the Nyquist frequency. The latter is thus avoided when
measuring the spectral index, while the former -- too small to be visible
in the spectra of individual light curve sections used to determine the 
evolution of $\gamma$ around the outburst cycle (see below) -- is ignored.
V344~Lyr has a much smaller spectral index and -- being fainter -- white
noise clearly takes over at high frequencies, limiting thus the useful
interval to measure the spectral index. $\gamma$, as measured in the average
power spectra, is listed in Table~2, separately for quiescent
and outburst phases.

\begin{table}
\label{Table: gamma}
	\centering
	\caption{Spectral index $\gamma$ measured in average power spectra of
V1504~Cyg, V344~Lyr during quiescence and outburst and KIC~9202990 (all data
combined), together with the frequency range used to measure $\gamma$.}

\begin{tabular}{lccc}
\hline

Object & frequency range (d$^{-1}$) & $\gamma$ (qui.) & $\gamma$ (outb.) \\
\hline
V1504 Cyg   & 35.5 \ldots 631 & 1.52 & 1.35 \\
V344 Lyr    & 50.1 \ldots 200 & 0.49 & 0.84 \\
KIC 9202990 & 15.8 \ldots 725 & \multicolumn{2}{c}{1.31} \\
\hline
\end{tabular}
\end{table}

\begin{figure*}
	\includegraphics[width=\textwidth]{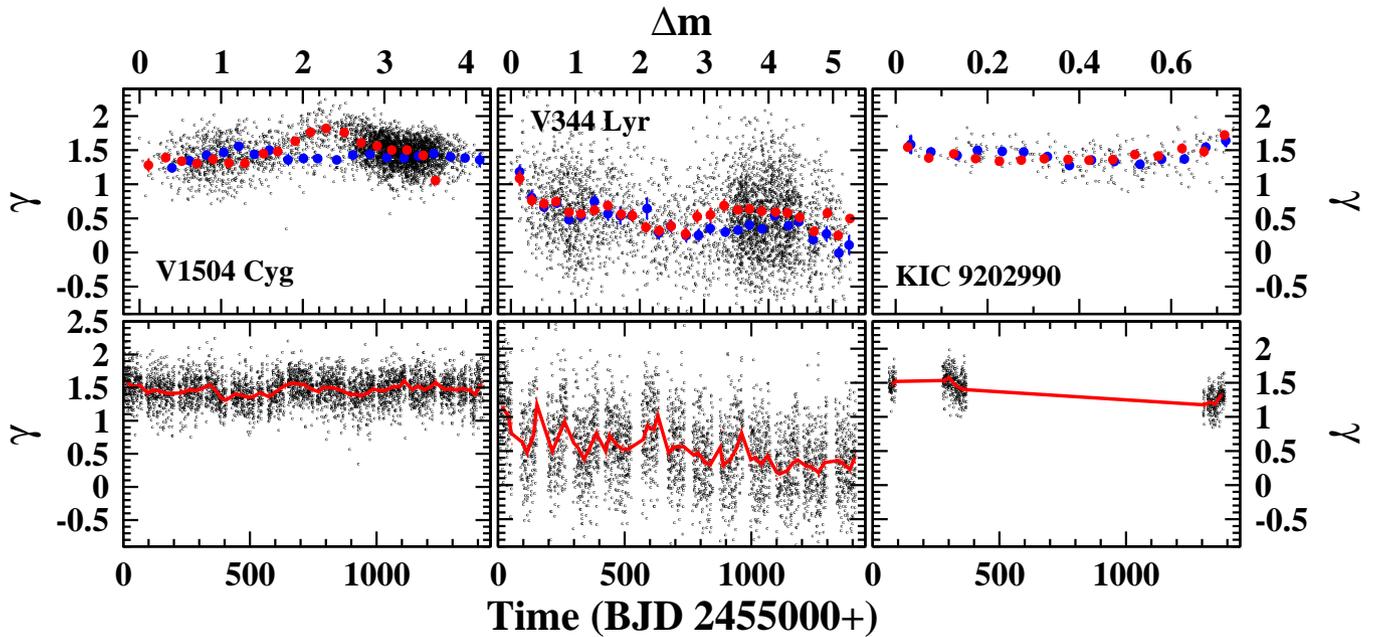}
      \caption[]{Spectral index $\gamma$ of the frequency distribution in
                 individual light curve sections (black dots) of three
                 systems, all drawn on the same scale. {\it Upper frames:} 
                 $\gamma$ as a function of the magnitude $\Delta m$ below
                 outburst maximum. The blue and red dots represent average
                 values for sections before and after maximum, respectively.
                 {\it Lower frames:} $\gamma$ as a function of time. The red
                 graphs connect average values in 25 day intervals.}
\label{spec-index}
\end{figure*}

In order to study variations of $\gamma$ around the outburst cycle,
Lomb-Scargle periodograms of the same light curve 
sections defined in Sect.~\ref{The strength of the flickering} were 
calculated after subtraction of variations on time scales longer
than 8~h. $\gamma$ 
was then measured as the slope of a linear least squares fit to the 
frequency intervals quoted in Table~2 of the periodograms 
in double logarithmic representation, binned in intervals of
$\Delta \log(f) = 0.05$. The results are shown in Fig.~\ref{spec-index}
where $\gamma$ is plotted as a function of the magnitude (using the same 
magnitude scale as in Sect.~\ref{The strength of the flickering}) in the upper
frames and against time in the lower ones. To facilitate a comparison,
the $y$-axis scale of the diagrams is the same for the different
systems. The blue (before maximum) and red (after maximum) dots 
in the upper graphs represent the average values 
of $\gamma$ in magnitude intervals of $\Delta m = 0.2$ (0.05 for KIC~9202990).
The error bars (often smaller than the plot symbols) 
indicate the standard error of the mean. In the lower frames, 
the red graphs connect the average values of $\gamma$ in time 
intervals of 25 days. 

For individual light curve sections the periodograms are not well enough
defined to measure $\gamma$ with high precision. Thus the
scatter is considerable, in particular for the fainter star V344~Lyr. 
But the ensemble of all individual values yields a systematic behaviour
which, however, is different for the three systems. 

V1504~Cyg is the most interesting case. No significant evolution of $\gamma$
occurs before and on the rise to outburst maximum. But during the decline
from outburst a significant increase is evident at intermediate magnitudes.
This is reminiscent of the hysteresis observed in the flickering amplitude of
the same system (as is the case in Fig~\ref{flickamp} the downward turn at 
the faint end of the red curve
is not considered significant as it is based on very few data 
points). Long term changes of $\gamma$ over time (lower left frame of \
Fig.~\ref{spec-index}) are not noticeable beyond small wiggles.

In contrast to V1504~Cyg, V344~Lyr, although much noisier, appears to 
exhibit a slight increase of $\gamma$ as the system approaches outburst
maximum. A marginal increase of $\gamma$ during decline is observed
just below $\Delta m = 2.8$. If real, this coincides with the magnitude range
where a slight hysteresis in the flickering amplitude is seen in this system.
As a function of time, $\gamma$ exhibit a small but systematic long term
decrease.

The amplitude of the variations of KIC~9202990 is much smaller than that
of the other two systems (the exceptionally low minima in the interval
JD~2456320--50 have been ignored). Similarly to the behaviour of the
flickering strength, there appears to be an 
upward turn of $\gamma$ close to the minimum and -- less clear -- close to 
the maximum of the mini-outbursts. No significant difference of the behaviour 
before and after the outbursts is observed. Concerning the long-term
evolution, a slight decrease of $\gamma$ is
observed in the last of the three distinct intervals 
(Fig.~\ref{kic9202990-longtlc}) during which KIC~9202990 was observed.

\section{Discussion}

The hysteresis of the flickering amplitude around the outburst cycle in
V1504~Cyg and V344~Lyr, the amplitude being
smaller during the rise to outburst than during decline, has never before 
been noticed in these or in any other system. This attests to the unparallel
quality and quantity of the Kepler data for the purpose of the current study. 
At the same time it calls for an explanation.

Considering the disk instability model (DIM) for dwarf novae outbursts
\citep[e.g., ][]{Lasota01}, the mere existence of a hysteresis can 
qualitatively be explained 
in a scenario where the flickering part of the flux of the accretion disk 
is a constant fraction of the locally emitted flux, but this fraction being 
higher/lower for those parts of the disk which are on the lower/upper branch 
of the famous S-shaped relation between the surface column density and the 
effective temperature. On the upper branch, the disk can be approximated by
a standard steady state accretion disk \citep{Shakura73}, and the local
temperature and thus the emitted flux is proportional to the local mass 
transfer rate and decreases strongly with increasing radius in the disk. 
The quiescent disk, on the other hand, is generally observed to have a much
smaller radial temperature gradient \citep[eg., ][]{Wood92, Vrielmann02, 
Baptista04}. Thus, during quiescence the surface brightness of the disk is 
not such 
a strong function of the distance from the central star and therefore the 
outer parts provide more flux to the total light compared to the inner parts 
than is the case during outburst maximum.
The observed shape of the outburst light curves of the currently
discussed systems (steep rise and more gradual decline) is normally
interpreted such that a heating front starts in the outer disk and 
then progates inwards, followed by a cooling front propagating in the same
direction \citep[e.g.][for the specific case of V1504~Cyg and V344 Lyr see
also \citet{Osaki13a} and \citet{Cannizzo10}]{Smak84, Cannizzo86}. At a 
given magnitude on the rise to the outburst the observed flickering is a
mixture of the flickering of the remaining quiescent inner part of the
disk and the lower flickering of the outer parts which are already in
outburst. At the same magnitude on the decline the situation is inversed. 
Since, however, the relative flux contribution of outer parts of a quiescent 
disk, as compared to the inner disk, is higher than in an outbursting
disk, stronger flickering is now seen. Thus, a hysteresis arises. 

In order to quantify this effect full DIM calculations and a realistic 
model for the radial distribution of flickering are required. This is
beyond the scope of this paper. However, a simple toy model can already provide
an approximation, demonstrating, if not the detailed properties, but at least 
the existence of a hysteresis. Both, quiescent and outbursting parts of the 
disk are taken to radiate like black bodies. During outburst the effective 
temperature is a function of the mass of the primary
star, the local mass transfer rate $\dot{M}$ (here assumed to be constant) and 
the distance from the centre of the disk \citep[eq.~5.43 of]{Frank02}. 
I assume the temperature during quiescence 
to decrease linearly form $T_{\rm i}$ at the inner disk rim to $T_{\rm o}$ 
at the outer rim. $T_{\rm i}$ and $T_{\rm o}$ are free parameters of the model.  
I simulate the outburst light curve by considering the disk to be 
composed of concentric rings which are either in quiescence or in outburst
and summing up their monochromatic
flux at the effective wavelength of 6650 \AA\ of the broad
Kepler bandpass \citep{Brown11} 
as a function of time. The velocity of the heating front its taken
to be $\alpha_{\rm h} c_{\rm s}$ \citep{Hameury20}, where $\alpha_{\rm h}$ is the
Shakura-Sunyaev viscosity parameter in the hot state, and $c_{\rm s}$ is the
local sound speed which can be calculated from Eqs.,~5.35, 5.26, and 5.49
of \citet{Frank02}. The cooling front is assumed to propagate more slowly 
by a constant factor 
of the heating front speed. But note that the front
speeds only determine the time scale of the outburst and have no bearing on
the flickering hysteresis. Transforming fluxes into
magnitudes, it is then possible to construct the dependence of the flickering
amplitude on the magnitude during both, outburst rise and decline.

The model has several parameters, but most of these have only a small
influence on the results. For the outer disk radius a typical value
of $0.4 a$ \citep{Warner95} is taken,
where $a$ is the component separation, determined by the stellar masses and
the orbital period. For the white dwarf mass I assume the average primary
star mass of $0.83\ M_\odot$ of CVs \citep{Zorotovic11}, and the mass ratio 
$q=0.16$ \citep{Osaki13b} and the period of V1504~Cyg. The white dwarf
radius according to \citet{Nauenberg72} defines the inner disk radius.
The eventual presence of (constant) third light (p.ex., the secondary star,
a hot spot at the disk rim, or the white dwarf) is neglected. The 
disk is assumed to be seen face on. 

\begin{figure}
	\includegraphics[width=\columnwidth]{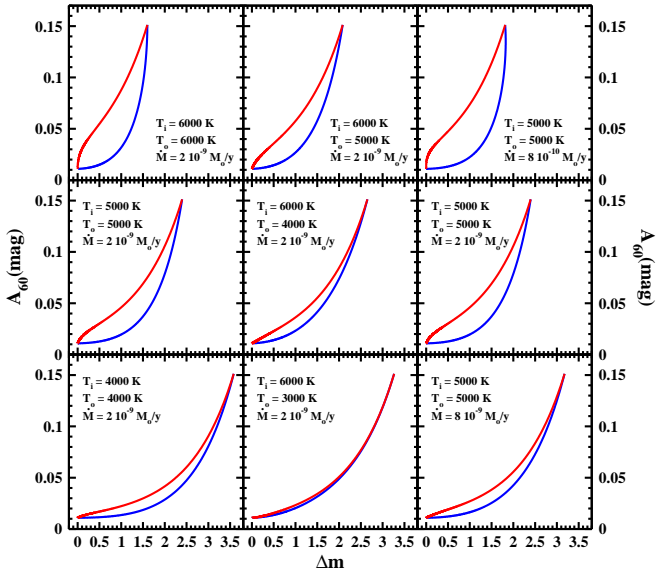}
      \caption[]{Results of simple model calculations aimed to reproduce the
                 hysteresis observed in the flickering amplitude $A_{60}$
                 during the
                 rise and decline of dwarf novae outbursts. In the various
                 frames the calculated flickering amplitude on the rise to
                 (blue) and the decline from (red) outburst is shown as a
                 function of the magnitude $\Delta m$ 
                 below outburst maximum for different
                 combinations of those model parameters which most strongly
                 determine the shape of the $\Delta m - A_{60}$ relation.
                 For details, see text.}
\label{simul}
\end{figure}

While this simple model is only capable to approximately replicate the 
outburst shape and falls somewhat short to simultaneously reproduce the
outburst amplitude, the hysteresis amplitude (i.e., the difference between the 
upper and the lower branch as a function of magnitude) and the magnitude
at the maximum of the hysteresis amplitude curve, Fig.~\ref{simul} shows
that, depending on the choice of parameters, the model can produce a 
sizable hysteresis of the magnitude -- flickering amplitude 
relation. It turns out that this curve depends mostly on $\dot{M}$, 
$T_{\rm i}$ and $T_{\rm o}$.

The left hand column of the figure shows this relation (i.e., 
$\Delta m$ vs.\ $A_{60}$)
for an assumed isothermal (probably unrealistic) quiescent disk with 
different temperatures, all for a constant mass transfer 
rate of $2 \times 10^{-9}\ {\rm M_\odot/y}$. As in Fig.~\ref{flickamp} 
the blue and red branches represent the rise to and decline from outburst, 
respectively. The fractional flickering flux in quiescence and outburst 
are free parameters of the model. They have been chosen such as to
roughly reproduce $A_{60}$ as observed during the respective states in 
V1504~Cyg. The hysteresis is strongly expressed if the disk temperature 
in quiescence is relatively high (upper left frame),
but then the outburst amplitude remains much lower than observed. A cooler
quiescent disk (4000~K) yields the observed outburst amplitude, but then the 
amplitude of the hysteresis is less than half of what is observed in 
V1504~Cyg (lower left frame). 
In the central column of the figure a temperature gradient in the quiescent 
disk is assumed, again fixing $\dot{M}$ to $2 \times 10^{-9} {\rm M_\odot/y}$. 
The hysteresis amplitude is then strongly dependent on $T_{\rm o}$ and 
vanishes when $T_{\rm o}$ becomes small. This is easily understood because then
the ratio of the surface brightness of inner and outer quiescent disk is not
much different from that of the disk in outburst. Finally the right hand
column shows the $\Delta m - A_{60}$ relation for an isothermal disk, assuming 
different values for $\dot{M}$. The outburst amplitude grows with increasing 
$\dot{M}$, but the hysteris amplitude decreases.  

Thus, this simple scenario, while not being able to reproduce all
features of the observed $\Delta m - A_{60}$ relation, is capable 
-- adopting reasonable parameters -- to explain the very existence of a 
hysteresis, {\it quod erat demonstrandum}. Details such as the observed 
convergence of the two branches at high magnitudes as well as the
hysteresis amplitude are not well modelled. Moreover, by design the model
cannot reproduce the flat part of the $\Delta m - A_{60}$ relation at high 
magnitudes, observed in all current objects except V1504~Cyg (disconsidering
the unusual system KIC~9202990). On the other
hand, selecting suitable parameters, the model has no problems to explain
the apparent absence of a hysteresis in V447~Lyr and V516~Lyr.

Turning the attention now to the frequency spectrum of the flickering and
particularly to details of the high frequency slope of the power spectra on
the double logarithmic scale, in past investigations the focus has often
been restricted to break points, i.e., frequencies at which the slope and thus
the spectral index exhibits a more or less sudden change
\citep[e.g.,][]{Dobrotka12, Dobrotka14, Dobrotka17, Scaringi12, Scaringi14}.
Various models have been put forward to reproduce the observed features in
a variety of different systems. They have been met with mixed success, 
indicating that none of them is sufficiently general to explain the variety of 
observational aspects of flickering. Moreover, it is not immediately obvious
how to interpret breaks in the power spectra and what determines specific
numerical values of the spectral index. 

Concerning the objects of the present study, \citet{Dobrotka15} investigated
the frequency spectrum of the same Kepler data of V1504~Cyg. A re-analysis 
of these data, together with an analysis of the Kepler data of V344~Lyr was
published by \citet{Dobrotka16}. In both stars they find a break at
$\log(f) = 1.5$ in quiescence\footnote{Note 
that \citet{Dobrotka16}
express frequencies in units of Hz, while I use units of $d^{-1}$ in this 
study.}. During outburst, this break persists, and additional breaks 
appear at $\log(f) = 1.9$ and between $2.1$ 
and $2$. \citet{Dobrotka17} conclude that the low 
flickering frequencies originate from regions 
affected by the outburst (the accretion disk or its inner edge) while the 
high frequencies are not affected and originate in an inner hot geometrically
thick corona. The authors suggest that flickering in both, 
quiescence and outburst, is generated by turbulent mass accretion, but
during outburst an additional non-turbulent radiation source must be
present. 

Comparing the present results with those of the previous studies, 
Fig.~\ref{powersp} indicates that the slope of the quiescent
power spectra of V1504~Cyg and V344~Lyr indeed changes close to
$\log(f) = 1.5$ to a degree that the range of lower 
frequencies was not used here to measure the spectral index. Additionally,
the high frequency regime of V1504~Cyg is not perfectly linear but
exhibits a slight hump in the approximate range  
$2.35 \le \log(f) \le 2.65$ which was also seen by 
\citet{Dobrotka15} (see their fig.~2; note the different definition of 
the $y$-axis in that figure). In contrast, I could not clearly identify
the additional break points claimed by \citet{Dobrotka16} to be present
during outburst.
     
It is not the purpose of this study to construct yet another model
in order to quantitatively reproduce the spectral index or the break 
points. I rather contend myself to investigate qualitatively whether
the index changes during the outburst cycle or over time. Thus, the exact
absolute value of the spectral index is less important that its variation.
This permits to ignore the slight deviations from linearity of the high 
frequency power spectrum slope which data noise inhibits anyway to be 
detected in the individual 8-hour sections of the light curves. 
The average spectral index over the approximately linear part of the
power spectra is therefore sufficient to characterize the frequency behaviour of
the flickering.

The long-term decrease of $\gamma$, seen in V344~Lyr and the slight
dependence of the average $\gamma$ on the epoch in KIC~9202990
(Fig.~\ref{spec-index}), may come as a surprise
because there is no apparent correlation with other changes in the long
term behaviour of the systems. Therefore, I just document this feature here,
without endeavouring to explain it.  

On the other hand, the systematic variations of $\gamma$ over the outburst 
cycle in V1504~Cyg and V344~Lyr indeed attest to changes 
of the flickering light source depending on the brightness and thus the state 
of the accretion disk. However, this dependence is not the same for the
two systems. While the average spectral index is, on average, slightly
smaller during outburst than in quiescence in V1504~Cyg 
[$1.371 \pm 0.010$ ($0 \le \Delta m \le 1.4$) vs.\ $1.467 \pm 0.004$
($2.6 \le \Delta m \le 4.2$); here, the eror is the standard error of the 
mean], there is a small
increase of $\gamma$ with increasing brightness in V344~Lyr. More interesting,
however, is a hysteresis similar to that observed for the flickering 
amplitude. On 
the rise to outburst (blue in Fig.~\ref{spec-index}) $\gamma$ is systematically
smaller than during decline (red) at intermediate magnitude levels. Again, this
effect is clearer in V1504~Cyg, but appears also to be real in V344~Lyr. 
As in the case of the flickering amplitude this may also have its 
explanation in systematic differences in the flickering
behaviour -- here the distribution of time scales of the instabilities 
leading to flickering -- 
in bright and faint parts of the accretion disk, together with
the different contributions of these parts during the outburst cycle.

\section{Summary}
\label{Summary}

In this study, I took advantage of the unparallel quantity and quality of 
high cadence data observed by the Kepler satellite of several dwarf novae 
to investigate some aspects of
flickering in these systems around their outburst cycles. In particular, 
the strength of the flickering and the spectral index of the high frequency
part of the flickering power spectrum were measured as a function of
magnitude. The main results can be summarized as follows:

(1) In three of four investigated dwarf nova the tendency, suspected earlier
by B21 based on data of lesser quality, for the flickering strength (on a
magnitude scale) to remain practically constant when the system is above a 
certain brightness threshold and to increase strongly at fainter magnitudes, 
is confirmed. 

(2) Two systems (V1504~Cyg and V344~Lyr) exhibit a clear hysteresis of the 
flickering strength around
the outburst cycle in the sense that flickering at a given brightness is 
weaker on the rise to outburst than during decline. In two other dwarf novae 
higher noise levels does not permit to verify if a similar hysteresis exists
or not. Qualitatively, the hysteresis can be explained in a DIM scenario for
dwarf nova outbursts with the flickering amplitude being lower/higher in parts 
of the accretion disk in a high/low state, and a smaller radial temperature 
gradient in the low state parts compared to the high state parts.

(3) The spectral index also varies around the outburst cycle, however, the
dependence on magnitude is not the same for the investigated systems. A
hysteresis similar to the one seen in the flickering strength is present
in V1504~Cyg and possibly also in V344~Lyr, the index being lower during
the rise to outburst than during decline at the same brightness level. This
can be explained in the same scenario as above, assuming that the frequency
distribution of the flickering is different in high and low state parts of 
the accretion disk.

(4) The flickering properties as a function of brightness of the unusual 
system KIC~9202990 which, instead of normal dwarf nova outburst exhibits a 
continuous series of small amplitude variations interpreted as stunted 
outbursts, are different from those of ``well behaved'' dwarf novae.

\section*{Acknowledgements}

This paper is exclusively based on data collected by the Kepler mission and 
obtained from the MAST data archive at the Space Telescope Science Institute 
(STScI). Funding for the Kepler mission is provided by the NASA Science 
Mission Directorate. STScI is operated by the Association of Universities 
for Research in Astronomy, Inc., under NASA contract NAS 5-26555.

\section*{Data availability}

All data used in the present study are publically available at the
Barbara A.\ Mikulski Archive for Space Telescopes 
(MAST):\\ https://mast.stsci.edu/portal/Mashub/clients/MAST/Portal.html.






\bsp	
\label{lastpage}
\end{document}